\documentclass[12pt]{article}
\usepackage{amssymb}
\usepackage{setspace}

\evensidemargin \oddsidemargin
\author{
Laure Gouba \\
\small The Abdus Salam International Centre for
Theoretical Physics (ICTP),\\
\small Strada Costiera 11,
I-34151 Trieste Italy.\\
\small Email: lgouba@ictp.it
\\\\
Domagoj Kovacevic \\
 \small University of Zagreb, Faculty of Electrical
Engineering and Computing, \\
\small Unska 3, 10000 Zagreb, Croatia.\\
\small Email: domagoj.kovacevic@fer.hr 
\\\\
Stjepan Meljanac\\
\small Ruder Boskovic Institute,
 Bijenicka c.54 HR-10002 Zagreb, Croatia.\\
\small Email: meljanac@irb.hr}

\begin{title}
{A General formulation  of the Moyal and Voros products and its physical 
interpretation}
\end{title}
\begin{document}
\maketitle
\abstract{A unifying perspective on the Moyal and Voros 
products and their physical mea\-nings has been recently 
presented in the literature, where the Voros formulation admits
a consistent physical interpretation. We define a star product $\star$,
in terms of an antisymmetric fixed matrix $\Theta$, and an arbitrary 
symmetric matrix $\Phi$, that is a generalization of the Moyal and 
the Voros products. We discuss the quantum mechanics and the physical meaning
of the generalized star product.}

\section{Introduction}
Quantum field theories on noncommutative spaces are an important
area of research in high energy physics due to the fact that they can be used 
as a tool to detect aspects of Planck scale physics, where one expects 
the spacetime to show noncommutative behavior, their emergence in string 
theory and also as a tool to regularize quantum field theories \cite{bal}. Studying 
quantum field on noncommutative spaces leads to better 
understanding of the structure and the setup of quantum field theory itself
\cite{rich}.
The starting point for a large part of what is now called the 
noncommutative geometry is the commutator
\begin{equation}\label{cr}
 x^\mu \star x^\nu - x^\nu \star x^\mu = i\theta^{\mu\nu},
\end{equation}
implemented via the Moyal product often written  in the asymptotic form
\begin{equation}
 f(x)\star_M g(x) = f(x)e^{\frac{i}{2}\theta^{ij}\stackrel{\leftarrow}{\partial}_i
\stackrel{\rightarrow}{\partial}_j} g(x),
\end{equation}
where $\theta^{ij}$ is a constant antisymmetric $2$-tensor.
It is a noncommutative, associative product introduced originally in 
quantum me\-cha\-nics. It comes from a Weyl map between functions and operators. 
The commutation relation (\ref{cr}) has been introduced in the spacetime 
context by Doplicher, Fredenhagen and Roberts \cite{dop}. The Moyal product is not the 
only product which gives the above commutation relation. There is also the 
Voros product. In fact, it has been shown that the two products can be cast 
in the same general framework in the sense that they are both coming from a ``Weyl map``.
More precisely, it has been shown that the Moyal product comes from a map, called 
the Weyl map, which associates operators to functions with symmetric ordering, while
the Voros one comes from a similar map, a weighted Weyl map, which associates 
operators to functions with normal ordering \cite{fed}. The Moyal and the Voros 
formulations of noncommutative field theory has been a point of controversy in the past.
This issue has been recently addressed in the context of noncommutative non relativitistic
quantum mechanics \cite{schol}. In particular, it has been shown that the two formulations 
simply correspond to two different representations associated with two different choices 
of basis on the quantum Hilbert space. The connection between the Voros and Moyal Weyl 
products has been shown in Ref.\cite{stern} and their equivalence is well known in the sense 
of Kontsevich \cite{kontse}. In the present paper, we define a star 
product in $2+ 1$ dimensional space-time, 
in terms of an antisymmetric fixed matrix $\Theta$ and an arbitrary symmetric matrix 
$\Phi$. This definition generalize the formulation of the Moyal and the Voros star products.
Our motivation is to explore the possible unification of the physical meaning of these 
star products. In section $2$, we define the generalized star product followed by the 
quantum mechanics associated to this star product in section $3$.
\section{Generalized star product}
We consider the $(2 + 1)$ dimensional space-time operators
where the operators $\{x_\mu\}_{\mu=0,1,2}$ satisfy the commutation relations
\begin{equation}\label{eqo}
 [x_\mu,\:x_\nu] = 0,\: \mu = 0,1,2,\:\nu = 0,1,2.
\end{equation}
We define, new operators
\begin{equation}\label{eq1}
 \hat x_\mu = x_\mu + \frac{i}{2}\Theta_{\mu\alpha}\partial_\alpha 
+\frac{i}{2}\Phi_{\mu\alpha}\partial_\alpha,
\end{equation}
where the matrix $\Theta$ is fixed antisymmetric and the matrix $\Phi$, 
an arbitrary symmetric matrix defined respectively as follows
 \begin{eqnarray}\label{mats}
 \Theta = \left(\begin{array}{lll}
0 & 0 & 0\\
0 & 0 & \theta \\
0 & -\theta & 0
\end{array}
\right),\quad
 \Phi = \left(\begin{array}{lll}
0 & 0 & 0\\
0 & \varphi_{11}& \varphi_{12} \\
0 & \varphi_{12} & \varphi_{22}
\end{array}
\right),
\end{eqnarray}
with the time taken to be an ordinary $c$-number.
The operators defined in equation (\ref{eq1}) satisfy the commutation relations
\begin{equation}\label{ncr}
 [\hat x_i,\:\hat x_j] = i\theta\epsilon_{i j},\: i = 1,2,\:j = 1,2\quad
\textrm{and}\quad
 [\hat x_0, \hat x_i] = 0,\: i = 1,2.
\end{equation}
We define a star product denoted by $\star$ as follows
\begin{equation}\label{pro}
 f(\mathbf{x}) \star g(\mathbf{x})
= (f\star g)(\mathbf{x}) = 
f(\mathbf{x})\exp(\frac{i}{2}(\Phi + \Theta)_{\mu\nu}
\stackrel{\leftarrow}{\partial}_\mu \stackrel{\rightarrow}{\partial}_\nu )
g(\mathbf{x}),
\end{equation}
that is associative but not commutative. 
From the general point of view, we define the commutator
of two functions with respect to the product (\ref{pro}) as
\begin{equation}\label{bra}
 [f,g]_{\star} = f \star g - g \star f,
\end{equation}
 that is bilinear, antisymmetric,
satisfies the Jacobi identity and the Leibniz rule
\footnote{$[f,\; g \star h]_\star = [f,\; g]_\star \star h + g \star [f,h]_\star$ }.
When $\Phi \equiv 0$, then the equation (\ref{pro}) is equivalent to the formulation of
the Moyal product $\star_{\textrm{M}}$ and when 
\begin{equation}\label{mvor}
 \Phi \equiv \Phi_\theta = \left(\begin{array}{lll}
               0 & 0 & 0\\
               0 &  -i\theta & 0\\
              0 & 0 & -i\theta
              \end{array}
		\right),
\end{equation}
the equation (\ref{pro}) is equivalent to the formulation of the 
Voros product $\star_{\textrm{V}}$.
Since the matrix $\Phi$ is arbitrary symmetric, it induces 
a family of star products that are all equivalent to the Moyal product. For the 
proof, 
we consider on the space of functions on the Minkowski space-time, 
where the metric is of signature $(-,+,+)$, the map
\begin{equation}
 T = e^{\frac{i}{4}\Phi_{\mu\nu}\partial_\mu\partial_\nu},\quad \mu = 0,1,2;\: \nu = 0,1,2
\end{equation}
and it is equivalence if 
\begin{equation}\label{equi}
 T(f \star_M g) = T(f)\star T(g).
\end{equation}
The convenient framework to show the equality (\ref{equi})
 is the momentum space, where in the momentum representation
\begin{equation}
 f(\bold{x}) = \int \frac{d^3 p}{(2\pi)^3}\tilde{f}(p)e^{i\bold{p}\cdot\bold{x}},
\end{equation}
with $\bold{p}\cdot\bold{x} = -p_0 x_0 + p_1x_1 + p_2x_2$.
We have
\begin{eqnarray}
 (f\star_M g)(\bold{x}) &=& \int\frac{d^3p}{(2\pi)^3}\frac{d^3q}{(2\pi)^3}\tilde{f}(p)
\tilde{g}(q)e^{-\frac{i}{2}\theta_{ij}p_iq_j}e^{i\bold{(p+q)}\cdot\bold{x}},
\end{eqnarray}
where 
\begin{equation}
 \theta_{ij}= \theta\epsilon^{ij},\quad i=1,2,\; j = 1,2.
\end{equation}
It is straightforward to show that
\begin{eqnarray}\nonumber
 T (f\star_M g)(\bold{x}) 
&=&  \int\frac{d^3 p}{(2\pi)^3}\frac{d^3 q}{(2\pi)^3}
\tilde{f}(p)\tilde{g}(q)e^{-\frac{i}{2}\theta_{ij}p_iq_j}\\\label{equ1}
&\times& e^{-\frac{i}{4}\left(\varphi_{11}(p_1 +q_1)^2 +\varphi_{22}(p_2 + q_2)^2
+ 2\varphi_{12}(p_1 + q_1)(p_2 +q_2)\right)}e^{i(\bold{p+q})\cdot\bold{x}}.
\end{eqnarray}
Now let us compute
\begin{eqnarray}\nonumber
 T(f) \star T(g) &=& e^{\frac{i}{4}\Phi_{ij}\partial_i\partial_j}\int\frac{d^3 p}{(2\pi)^3}
\tilde{f}(p)e^{i\bold{p}\cdot\bold{x}}\:\star \:
e^{\frac{i}{4}\Phi_{ij}\partial_i\partial_j}\int\frac{d^3 q}{(2\pi)^3}
\tilde{g}(q)e^{i\bold{q}\cdot\bold{x}}\\\nonumber
 &=& \int\frac{d^3p}{(2\pi)^3}\frac{d^3q}{(2\pi)^3}\tilde{f}(p)
\tilde{g}(q)
e^{\frac{-1}{4}(\varphi_{11}(p_1^2 +q_1^2) 
+ \varphi_{22}(p_2^2 +q_2^2) + 2\varphi_{12}(p_1 p_2 + q_1q_2)}\\\label{yes}
&\times& e^{i\bold{p}\cdot\bold{x}}\:
e^{\frac{i}{2}(\Phi +\Theta)_{ij}\stackrel{\leftarrow}{\partial}_i
\stackrel{\rightarrow}{\partial}_j}\: e^{i\bold{q}\cdot\bold{x}}.
\end{eqnarray}
Since
\begin{eqnarray}\label{oui}
  e^{i\bold{p}\cdot \bold{x}}\:
e^{\frac{i}{2}(\Phi +\Theta)_{ij}\stackrel{\leftarrow}{\partial}_i
\stackrel{\rightarrow}{\partial}_j}\: e^{i\bold{q}\cdot\bold{x}} =
e^{ -\frac{i}{2}\theta_{ij}p_iq_j }
e^{\frac{-i}{2}(\varphi_{11}p_1q_1 +\varphi_{22}p_2q_2 +
\varphi_{12}(p_1q_2 + p_2 q_1))} e^{i(\bold{p+q})\cdot\bold{x}}.
\end{eqnarray}
The equation (\ref{yes}) transforms to 
\begin{eqnarray}\nonumber
  T(f) \star T(g) &=&
\int\frac{d^3 p}{(2\pi)^3}\frac{d^3 q}{(2\pi)^3}
\tilde{f}(p)\tilde{g}(q)e^{-\frac{i}{2}\theta_{ij}p_iq_j}\\\label{equ2}
&\times& e^{-\frac{i}{4}\left(\varphi_{11}(p_1 +q_1)^2 +\varphi_{22}(p_2 + q_2)^2
+ 2\varphi_{12}(p_1 + q_1)(p_2 +q_2)\right)}e^{i(\bold{p+q})\cdot\bold{x}}.
\end{eqnarray}
The following equality 
\begin{equation}
  T (f\star_M g) =  T(f) \star T(g),
\end{equation}
holds for any symmetric matrix $\Phi$.
\section{Quantum Mechanics with star product}
In this section, we consider the formalism of noncommutative quantum 
mechanics as in Ref. \cite{laure}, where noncommutative quantum mechanics is
 formulated as a quantum system on the Hilbert space of Hilbert-Schmidt operators
acting on classical configuration space.
Here, we consider $(2 + 1)$ dimensional space-time with only spacial noncommutativity.
Restricting to two dimensions, the coordinates of the noncommutative configuration
space satisfy the commutation relation
\begin{equation}
 [\hat x_1,\:\hat x_2] = i\theta.
\end{equation}
It is convenient to define the creation and annihilation operators 
\begin{equation}
 b = \frac{1}{\sqrt{2\theta}}(\hat x_1 + i\hat x_2);\quad
b^\dagger = \frac{1}{\sqrt{2\theta}}(\hat x_1 - i\hat x_2),
\end{equation}
where $\dagger$ is the hermitian conjugate notation,
that satisfy the Fock algebra $[b,\; b^\dagger] = 1$. 
The noncommutative configuration space
is then isomorphic to boson Fock space
\begin{equation}
 \mathcal{H}_c = \textrm{span}\left\{ |n\rangle = \frac{1}{\sqrt{n!}}(b^\dagger)^n|0\rangle
\right\}_{n=0}^{n=\infty},
\end{equation}
where the span is taken over the field of complex numbers.

The quantum Hilbert space, is identified with the set of Hilbert Schmidt operators acting 
on noncommutative configuration space
\begin{equation}\label{hilb}
 \mathcal{H}_q = \{\psi(\hat x_1,\hat x_2) :\psi(\hat x_1,\hat x_2)\in 
\mathcal{B}(\mathcal{H}_c), tr_c[\psi^\dagger(\hat x_1,\hat x_2)\psi(\hat x_1,\hat x_2)] < \infty \}.
\end{equation}
Here $tr_c$ denotes the trace over noncommutative configuration space and 
$\mathcal{B}(\mathcal{H}_c)$ the set of bounded operators on $\mathcal{H}_c$.
This space has a natural inner product and norm 
\begin{equation}\label{inn}
 (\phi(\hat x_1,\hat x_2),\;\psi(\hat x_1,\hat x_2)) = 
tr_c[\phi^\dagger(\hat x_1,\hat x_2)\psi(\hat x_1,\hat x_2)],
\end{equation}
and form a Hilbert Space. 
To distinguish the noncommutative configuration space, which is also
a Hilbert space, from the quantum Hilbert space above, we use the 
notation $|\cdot \rangle$ for elements of the noncommutative configuration
space, while elements of the quantum Hilbert space are denoted by 
$\psi(\hat x_1,\;\hat x_2)\equiv |\psi)$. The elements of its dual 
(linear functionals) are as usual denoted by bras, $(\psi |$, which maps
elements of $\mathcal{H}_q$ onto complex numbers by 
$(\phi|\psi) = (\phi,\psi) = tr_c[\phi^\dagger(\hat x_1,\hat x_2)\psi(\hat x_1,\hat x_2)].$
We reserve the notation $\dagger$ for Hermitian 
conjugation on noncommutative configuration space and the notation $\ddagger$ 
for Hermitian conjugation on quantum Hilbert space. 
The operators acting on the quantum Hilbert space are denoted by capital letters.
The noncommutative Heisenberg algebra in two dimensions 
\begin{eqnarray}
\left[\hat x_1,\;\hat x_2\right] = i\theta,\:
\left[\hat x_1, \hat{p}_1\right] = i\hbar,\:
\left[\hat x_2, \hat{p}_2\right] = i\hbar,\:
\left[\hat{p}_1,\;\hat{p}_2\right] = 0,\:
\left[\hat x_1,\;\hat{p}_2\right] = 0,\:
\left[\hat x_2,\;\hat{p}_1\right] = 0,
\end{eqnarray}
is now represented in terms of operators $\hat X_1,\:\hat X_2$ and 
$\hat P_1,\: \hat P_2$ acting on the quantum Hilbert space (\ref{hilb})
with the inner product (\ref{inn}), which is the analog of the Schr\"odinger representation of
the Heisenberg algebra. These operators are given by 
\begin{equation}
 \hat X_1\psi(\hat x_1,\;\hat x_2) = \hat x_1\psi(\hat x_1,\:\hat x_2),\quad
\hat X_2\psi(\hat x_1,\;\hat x_2) = \hat x_2\psi(\hat x_1,\;\hat x_2),
\end{equation}
\begin{equation}
 \hat P_1\psi(\hat x_1,\:\hat x_2) = 
\frac{\hbar}{\theta}[\hat x_2,\;\psi(\hat x_1,\hat x_2)],\quad
\hat P_2\psi(\hat x_1,\:\hat x_2) = 
-\frac{\hbar}{\theta}[\hat x_1,\;\psi(\hat x_1,\hat x_2)].
\end{equation}
The position operators act by left multiplication and the momentum 
acts adjointly. It is also useful to introduce the following quantum 
operators
\begin{equation}
 B = \frac{1}{\sqrt{2\theta}}(\hat X_1 + i\hat X_2),\quad
B^\ddagger = \frac{1}{\sqrt{2\theta}}(\hat X_1 -i \hat X_2),\quad
\hat P = \hat P_1 + i\hat P_2,\quad
\hat P^\ddagger = \hat P_1 - i\hat P_2.
\end{equation}
These operators act as following
\begin{eqnarray}\nonumber
 B\psi(\hat x_1,\hat x_2) &=& b \psi(\hat x_1,\hat x_2),\quad
B^\ddagger\psi(\hat x_1,\hat x_2) = b^\dagger\psi(\hat x_1,\hat x_2),\\
\hat{P}\psi(\hat x_1,\hat x_2) &=& -i\hbar\sqrt{\frac{2}{\theta}}[b,\;
\psi(\hat x_1,\hat x_2)],\quad
\hat{P}^\ddagger\psi(\hat x_1,\hat x_2) = i\hbar\sqrt{\frac{2}{\theta}}[b^\dagger,
\;\psi(\hat x_1,\hat x_2)].
\end{eqnarray}
Let us consider now the above formalism in a system of units such that $\hbar = 1$.
The momentum eigenstates $|p)$ are given by
\begin{equation}
 |p) = \sqrt{\frac{\theta}{2\pi}} e^{ip\cdot \hat{x}},\quad
\hat P_i|p) = p_i|p),
\end{equation}
and they satisfy the usual resolution of identity and orthogonality condition
\begin{equation}\label{mpro}
 \int d^2 p |p)(p| = Id, \quad (p|p') = \delta(p_1-p_1^{'})\delta(p_2 -p_2^{'}).
\end{equation}
We consider now the following states as in \cite{schol} obtained by expansion
in terms of the momentum states as follows
\begin{equation}
 |x) = \int\frac{d^2 p}{2\pi}e^{-i\bold{p}\cdot\bold{x}}|p).
\end{equation}
We have
\begin{equation}
 (p|x) = \frac{1}{2\pi}e^{-i\bold{p}\cdot\bold{x}},\quad \quad (x|x') = \delta(x_1-x_1^{'})
\delta(x_2 -x_2^{'}),
\end{equation}
that means that the states $|x)$ are orthogonal.
Do these states resolve the identity with respect to the star product $\star$ as follows
\begin{equation}\label{mid}
 \int d^2 x|x)\star (x| = Id\: \textrm{?}
\end{equation}
In order to respond to this question, we 
compute
\begin{eqnarray}\nonumber
(p|\left(\int d^2 x|x)\star(x|\right)|p')
&=& 
e^{\frac{i}{2}\left(\varphi_{11}p_1p_1^{'} + (\varphi_{12}+\theta)p_1p_2^{'}+
(\varphi_{21}-\theta)p_2p_1^{'} +\varphi_{22}p_2p_2^{'}  \right)}
e^{\frac{i\theta}{2}(p_1p_2 +p_1^{'}p_2^{'})}\\\label{mai}
&\times& e^{-i\theta p_2p_1^{'}}\delta(p_1-p_1^{'})\delta(p_2 -p_2^{'}).
\end{eqnarray} 
Setting the matrix $\Phi \equiv 0$, then 
\begin{eqnarray}\nonumber
 (p|\left(\int d^2 x|x)\star_M(x|\right)|p') &=& e^{\frac{i\theta}{2}(p_1p_2^{'}- p_2p_1^{'}
+ p_1p_2 +p_1'p_2'-2p_2p_1')}\delta(p_1-p_1')\delta(p_2 -p_2')\\
 &=& (p|p').
\end{eqnarray}
In general, 
\begin{equation}
 (p|\left(\int d^2 x|x)\star(x|\right)|p') \neq (p|p').
\end{equation}
The states $|x)$ are then orthogonal
and resolve the identity with respect to the Moyal product, then constitute a basis of
the Hilbert space. Although this provides a consistent interpretational framework, the 
measurement of position needs more careful consideration as  the position operators
do not commute and thus a precise measurement of one of these observables leads to total
uncertainty in the other. In order to preserve the notion of position in the sense of a
particle being localized around a certain point, the best is to construct a minimum uncertainty
state in noncommutative configuration space and use that to give meaning to the notion of position.

The minimum uncertainty states on noncommutative configuration space, which are isomorphic
to the boson Fock space, are well known to be the normalized coherent states
\begin{equation}
 | z \rangle = e^{-z\bar{z}/2}e^{zb^\dagger}|0\rangle,
\end{equation}
where $ z = \frac{1}{\sqrt{2\theta}}(x_1+ix_2)$ is a dimensionless complex number
that satisfies the relation $b|z \rangle = z |z\rangle$. 
These states provide an overcomplete basis on the noncommutative configuration space.
Corresponding to these states, we can construct 
a state (operator) in quantum Hilbert space as follows
\begin{equation}
 | z,\bar{z}) = |z\rangle\langle z|,
\end{equation}
and these states satisfy 
\begin{equation}
 B|z,\bar{z}) = z |z,\bar{z}),
\end{equation}
and
\begin{equation}
 (z',\bar{z}'|z,\bar{z}) = tr_c[(|z'\rangle\langle z'|)^\dagger(|z\rangle\langle z|)]
= e^{-|z- z'|^2}.
\end{equation}
The star product defined in equation (\ref{pro}) can be expressed in terms
of complex variables as
\begin{eqnarray}\label{esp}
 \star \equiv
 e^{\frac{i}{4\theta}
\left[
(\varphi_{11}-\varphi_{22} + 2i\varphi_{12})\stackrel
{\leftarrow}{\partial_z}\stackrel
{\rightarrow}{\partial_{z}}
 +
(\varphi_{11}+\varphi_{22} - 2i\theta)\stackrel
{\leftarrow}{\partial_z}\stackrel
{\rightarrow}{\partial_{\bar{z}}}
+
(\varphi_{11}+\varphi_{22} + 2i\theta)\stackrel
{\leftarrow}{\partial_{\bar{z}}}\stackrel
{\rightarrow}{\partial_{z}}
+
(\varphi_{11}-\varphi_{22} -2i\varphi_{12})\stackrel
{\leftarrow}{\partial_{\bar{z}}}\stackrel
{\rightarrow}{\partial_{\bar{z}}}
\right]} .
\end{eqnarray}
When $\Phi\equiv 0$, we recognize the form of the Moyal product
\begin{equation}
 f(z,\bar{z})\star g(z,\bar{z}) =f(z,\bar{z}) e^{\frac{1}{2}(\stackrel
{\leftarrow}{\partial_z}\stackrel
{\rightarrow}{\partial_{\bar{z}}}-\stackrel
{\leftarrow}{\partial_{\bar{z}}}\stackrel
{\rightarrow}{\partial_{z}})}g(z,\bar{z}) = f(z,\bar{z})\star_M g(z,\bar{z}),
\end{equation}
and for non trivial matrix $\Phi_\theta$  defined in equation (\ref{mvor}), we 
recognize the form of the Voros product,
\begin{equation}
 f(z,\bar{z})\star g(z,\bar{z}) =f(z,\bar{z}) e^{\stackrel
{\leftarrow}{\partial_z}\stackrel
{\rightarrow}{\partial_{\bar{z}}}}g(z,\bar{z}) = f(z,\bar{z})\star_V g(z,\bar{z}).
\end{equation}
The question is whether the states $|z,\bar{z})$ resolve the identity 
\begin{equation}\label{resi}
 \int\frac{dzd\bar{z}}{\pi}|z,\bar{z})\star (z,\bar{z}| = Id\:\textrm{?}
\end{equation}
We now introduce the momentum eigenstates 
\begin{equation}
 |p) = \sqrt{\frac{\theta}{2\pi}}e^{i\sqrt{\frac{\theta}{2}}(\bar{p}b + p b^\dagger)},
\quad \int d^2 p|p)(p| = Id,\quad \hat{P}_i|p) = p_i|p),
\end{equation}
normalised such that $(p|p') = \delta(p_1-p_1')\delta(p_2-p_2')$.
The overlap of this basis with the momentum eigenstate is given by
\begin{equation}\label{mom}
 (z,\bar{z}|p) = \sqrt{\frac{\theta}{2\pi}}
e^{-\frac{\theta|p|^2}{4}}e^{i\sqrt{\frac{\theta}{2}}(p\bar{z} +\bar{p}z)}.
\end{equation}
We can check if equation (\ref{resi}) is satisfied with respect to the star product
$\star$ as follows
\begin{eqnarray}\nonumber
 \int\frac{dzd\bar{z}}{\pi}(p'|z,\bar{z})\star(z,\bar{z}|p) 
&=& e^{\frac{-\theta}{4}(|p|^2+|p'|^2)}
e^{\frac{i}{8} \left[
(\varphi_{11}-\varphi_{22} + 2i\varphi_{12})\bar{p}'\bar{p}
 +
(\varphi_{11}+\varphi_{22} - 2i\theta)\bar{p}p'\right]}\\\nonumber
&\times&
e^{\frac{i}{8} \left[
(\varphi_{11}+\varphi_{22} + 2i\theta)p\bar{p}'
+
(\varphi_{11}-\varphi_{22} -2i\varphi_{12})p'p
\right]}\delta(p_1-p_1')\delta(p_2-p_2').\\\label{hope}
\end{eqnarray}
For the particular case of the matrix $\Phi_\theta$ defined in (\ref{mvor}),
the equation (\ref{hope}) becomes 
\begin{equation}
 \int\frac{dzd\bar{z}}{\pi}(p'|z,\bar{z})\star_V(z,\bar{z}|p) 
= e^{\frac{-\theta}{4}(|p|^2+|p'|^2)}e^{\frac{\theta}{2}\bar{p}p'}
\delta(p_1 -p_1')\delta(p_2-p_2')
= (p|p'),
\end{equation}
that implies the resolution of the identity
\begin{equation}\label{vp}
 \int\frac{dzd\bar{z}}{\pi}|z,\bar{z})\star_V (z,\bar{z}| = Id.
\end{equation}
In general 
\begin{equation}
 \int\frac{dzd\bar{z}}{\pi}(p'|z,\bar{z})\star(z,\bar{z}|p)
\neq (p'|p).
\end{equation}
This means that the states $|z,\bar{z})$ do not resolve the identity
operator with respect to the star product $\star$.
\section{Conclusion}
We have defined a star product $\star$ in terms of a fixed antisymmetric 
matrix $\Theta$ and an arbitrary symmetric matrix $\Phi$. We have shown 
that the Moyal and the Voros products are some particular cases of the 
star product $\star$. 
This formulation confirms their equivalence
from a mathematical perpective \cite{schol}. 
As the matrix $\Phi$ is arbitrary, the star product $\star$ induces a family 
of star products with respect to $\Phi$ that are all equivalent to the 
Moyal product. In order to interpret the physical 
meaning of the star product, we set the problem in a completely general and 
abstract operator formulation of non-commutative quantum field theory and 
quantum mechanics. We expect to have some results that unify at least 
the physical meaning of the Moyal and the Voros products and then a complete
generalization of the results in \cite{schol}.
We set some physical states $|x)$ as expansion of the momentum states and they do 
resolve the identity with respect to the Moyal product. In the coherent states framework, 
the states $|z,\bar{z})$ resolve the identity with respect to the Voros product.
 For both of the states $|x)$ and $|z,\bar{z})$, we could not conclude the resolution of 
identity with respect to the star product $\star$. The physical interpretation of the 
star product $\star$ could not be pursued. 
Restricting to the Moyal and the Voros star products, in \cite{schol}, it has been
shown that only the Voros product can be interpreted as describing
a maximally localized system. That is also reflected in the transition amplitudes 
that differ with only the Voros amplitude representing the physical amplitude.
Similarly, in \cite{gor, gorb}, it has been shown that the low energy dynamics in 
the lowest Landau level approximation in relativistic quantum field theories in 
a magnetic field is described by the Voros rather than the Moyal product. 
It is known that a set of states that is 
overcomplete without having a resolution of identity is not practically useful. So
for the  star product $\star$ to be practically useful, one has to redefine an overcomplete 
set of states that resolve the identity with respect to this star product,
 alternatively, one should set the matrix $\Theta$ and the matrix $\Phi$ 
in a more general way. This setting may turn out to be more complicated, 
 but one may discover other star products equivalent to the Moyal product that 
unify the physical meaning of the Moyal and the Voros products. 

{\bf Acknowledgments}

The work of LG is supported by the Associate and Federation Scheme and the High
Energy Section of ICTP. LG would like to thank the Ruder Boskovic Institute for
her visit during which the work has been started. DK and SM were supported by the 
Ministry of Science and Technology of the Republic of Croatia under contract 
No. $098\textrm{-}0000000\textrm{-}2865$.


\begin{thebibliography}{20}
\bibitem{bal}{A. P. Balachandran, {\it Quantum Spacetimes in the Year $1$},
 arXiv: hep-th/$0203259$}.
\bibitem{rich}{R. J. Szabo, Phys. Rep. $378(2003)$ $\bf{207}$, arXiv:hep-th/$0109162$}.
\bibitem{dop}{S. Doplicher, K. Fredenhagen and J. E. Roberts, Phys. 
Lett. $\bold{B}\;331,\; 39 (1994)$.}
\bibitem{fed}{Salvatore Galluccio, Fedele Lizzi and Patrizia Vitale, Phys. Rev.
$\bf{D\; 78}$, $085007(2008)$}.
\bibitem{schol}{ Prasad Basu, Biswajit Chakraborty and
 Frederik G. Scholtz, {\it A unifying perspective on the Moyal and Voros products and
their physical meaning}, $2011$ J. Phys. A: Theor. $44 285204$}.
\bibitem{stern}{A. Pinzul, A. Stern, Int. J. Mod. Phys. A$20\;(2005)\; 5871-5890$ 
hep-th/$0406068$}.
\bibitem{kontse}{Maxim Kontsevich, {\it Deformation Quantization of Poisson Manifold, I.}
Lett. Math. Phys. $66,\; 157 - 216,\; 2003$; q-alg/$9709040$}.
\bibitem{laure}{F. G. Scholtz, L. Gouba, A. Hafver, C. M. Rohwer, J. Phys.
$\bf{ A\; 42}$ $2009,\; 175303$}.
\bibitem{gor}{Gorbar E. V. and Miransky V. A. $2004$ Phys. Rev. $\bold{D\; 70}$ $105007$}.
\bibitem{gorb}{Gorbar E. V., Homayouni S. and Miransky V. A. $2005$ Phys. Rev. 
$\bold{D\; 72}$ $065014$}.

\end{thebibliography}
\end{document}